\documentclass[prl,aps,twocolumn,superscriptaddress,showpacs,floatfix]{revtex4}
\usepackage{graphicx}
\usepackage{amsmath}

\begin{document}

\title{Anomalous Curie response of impurities \\
in quantum-critical spin-1/2 Heisenberg antiferromagnets}

\author{Kaj H. H\"oglund} 
\affiliation{Department of Physics, {\AA}bo Akademi University, 
Porthansgatan 3, FI-20500, Turku, Finland}

\author{Anders W. Sandvik} 
\affiliation{Department of Physics, Boston University, 590 Commonwealth Avenue,
Boston, Massachusetts 02215}

\date{\today}


\begin{abstract}
We consider a magnetic impurity in two different $S=1/2$ Heisenberg bilayer antiferromagnets at 
their respective critical inter-layer couplings separating N\'eel and disordered ground states. 
We calculate the impurity susceptibility using a quantum Monte Carlo method. With intra-layer 
couplings in only one of the layers (Kondo lattice), we observe an anomalous Curie constant 
$C^*$, as predicted on the basis of field-theoretical work [S. Sachdev et al., Science {\bf 286}, 
2479 (1999)]. The value $C^*=0.262 \pm 0.002$ is larger than the normal Curie constant $C=S(S+1)/3$.
Our low-temperature results for a symmetric bilayer are consistent with a universal $C^*$.
\end{abstract}

\maketitle

Investigating effects of various types of impurities is a promising strategy for studying
the electronic structure of strongly correlated systems \cite{mahajan,takigawa}. In very small 
concentrations, detectable changes in response functions cuaused by the impurities probe the 
inherent bulk properties of the host material and can give importants insights complementing
information derived from direct studies of the bulk. 
This strategy was recently examined in a field-theoretical 
study \cite{sachdev1,sachdev2} of a nearly quantum-critical two-dimensional (2D) antiferromagnetic 
host system with a single localized spin-$S$ impurity. Detailed predictions for the magnetic 
response of the impurity were made in the different finite-temperature scaling regimes associated 
with the quantum phase transition occuring in the host (as a function of some coupling constant 
$g$); from a quantum disordered spin-gapped paramagnet ($g > g_c$) to a gapless antiferromagnet 
($g < g_c$). The $T\to 0$ asymptotic impurity susceptibility---defined as the difference between 
the susceptibilities of systems with and without impurity---was predicted to have a Curie form, 
$\chi^z_{\rm imp} \to C/T$, with the Curie constant $C$ taking different values depending on 
the coupling \cite{sachdev1};
\begin{align}
C &={S^2}/3, & g&<g_c,\label{eq:rcsusc} \\
C & = C^* = {\tilde{S}(\tilde{S}+1)}/3,~~ \tilde S \not= S, & g&=g_c,\label{eq:qcsusc}\\
C & ={S(S+1)}/3, &g&>g_c \label{eq:qdsusc},
\end{align}
where we have set $\hbar =k_{B}=1$.
The classical-like response for $g < g_c$, which can be understood as being due to the impurity 
spin aligning with a large cluster of correlated spins (exponentially divergent, thus with essentially 
classical dynamics as $T \to 0$), was confirmed 
in a recent numerical study~\cite{khh1}, however only up to a logarithmic correction [formally
resulting in a log-divergent $C(T)$]. The log-correction was subsequently 
derived quantitatively in a different field theoretical formulation \cite{sachdev2} 
(and qualitatively also using spin wave theory \cite{sushkov}). It is given in terms of known 
ground-state constants of the host and is in complete agreement with the numerical results
\cite{khh2}. In the paramagnetic phase the host response is exponentially small due to the spin 
gap, and the usual Curie prefactor $S(S+1)/3$ is due solely to the localized impurity moment. 

The most remarkable prediction of Ref.~\cite{sachdev1} is the anomalous Curie constant (\ref{eq:qcsusc}) 
in the quantum critical regime, i.e., for $T < \rho_s,\Delta$, where $\rho_s$ is the spin 
stiffness in the N\'eel phase and the spin gap in the paramagnetic phase \cite{chn} (both of 
which vanish continuously at $g_c$). It was suggested that $S^2/3< C^* <S(S+1)/3$ and that this 
could be interpreted in terms of a fractional impurity spin; $\tilde S \not= S$. However, the 
anomalous Curie constant was challenged by Sushkov \cite{sushkov0}, on the basis of a Green's 
function theory giving $\tilde{S}=S$ also in the quantum critical regime \cite{note0}. The only 
apparent way to settle this issue is by explicit unbiased numerical computation of the 
impurity susceptibility of a quantum-critical system. To our knowledge, the only attempt so far is 
by Troyer, who carried out a quantum Monte Carlo (QMC) study of the $S=1/2$ Heisenberg bilayer with 
a vacancy (effectively corresponding to an $S=1/2$ impurity) \cite{troyer}. This calculation yelded
$\tilde S = S$ within statistical errors of a few percent, and thus, based on this study, the 
anomaly is either not present or is very small.

In this Letter we present results of large-scale numerical efforts to resolve the controversy
over the existence of an effectively  fractional impurity spin at the quantum-critical point.
We also  obtain further insights into the temperature dependence of the impurity susceptibility 
(corrections to the Curie form). We will present evidence of an anomalous Curie constant in the
case of an $S=1/2$ impurity. Our result is $C^* = 0.262 \pm 0.002$, which falls outside the
range $S^2/3< C^* <S(S+1)/3$. It should be noted, however, that the 
$\epsilon$-expansion presented for $C^*$ in Ref.~\cite{sachdev1} was not evaluated 
explicitly---the actual value was only conjectured to fall within the above range. Thus there 
is no contradiction. For an $S=1$ impurity we cannot detect  an anomaly within statistical 
errors, indicating that $C^*$ approaches $S(S+1)/3$ with increasing $S$.

\begin{figure}
\includegraphics[width=3.5cm,clip]{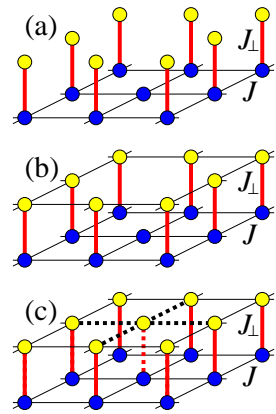}
\caption{(Color online) Spin impurity models. The circles represent $S=1/2$ spins
with nearest-neighbor interactions $J$ (in-plane) and $J_\perp$ (inter-plane). A vacancy 
in the incomplete bilayer (a) or full bilayer (b) host constitutes, effetively, an $S=1/2$ impurity. In
(c), changing the signs ($J\to -J$ and $J_{\perp}\to -J_{\perp}$) of the interactions
(the five dotted bonds) of one of the spins results in an $S=1$ impurity.}
\label{fig:fig1}
\end{figure}

We use an efficient stochastic series expansion (SSE) QMC 
technique \cite{sse} to compute the susceptibility of a single static $S=1/2$ or $S=1$ impurity 
in a quantum critical host. We consider two different host systems---a bilayer with intra- and 
inter-plane couplings $J$ and $J_\perp$, respectively, and an "incomplete bilayer" in which the 
$J$-coupling is present only in one of the layers (a Kondo lattice \cite{doniach}). The lattices 
and the ways in which we introduce the impurities in them are illustrated in Fig.~\ref{fig:fig1}. 
The Heisenberg hamiltonian for both host systems can be written as
\begin{eqnarray}\label{eq:hamiltonian}
H=&J&\sum _{\left <i,j\right >}\mathbf{S}_{1,i}\cdot
\mathbf{S}_{1,j}+\lambda J\sum _{\left <i,j\right >}\mathbf{S}_{2,i}\cdot
\mathbf{S}_{2,j}\nonumber \\
&+&J_{\perp}\sum _{i}\mathbf{S}_{1,i}\cdot \mathbf{S}_{2,i},
\end{eqnarray}
where $\left <i,j\right >$ denotes a pair of nearest neighbors on a periodic square lattice 
with $L\times L\times 2$ sites
and $\mathbf{S}_{a,i}$ is the usual spin-$\frac{1}{2}$
operator at site $i$ on layer $a=1,2$. We define the tuning parameter as $g=J_\perp /J$, and
$J,J_\perp > 0$. The symmetric and incomplete bilayers correspond to $\lambda =1$ and
$\lambda =0$, respectively. The quantum-critical points for these models have recently been 
extracted to high precision; $g_c= 2.5220(3)$ and $g_c=1.3888(1)$ for the full 
and icomple bilayers, respectively \cite{wang}. The ground state is N\'eel ordered for $g < g_c$ 
and disordered (spin-gapped) for $g>g_c$. As shown in Fig.~\ref{fig:fig1}, by removing a single 
spin from the incomplete (a) and symmetric (b) bilayers we effectively introduce $S=1/2$
impurities, as the remaining spin in the opposite layer is unpaired. An $S=1$ impurity (c) 
is obtained by making the five bonds to a given spin ferromagnetic (we here keep the magnitudes 
of these interactions unchanged).

Our investigations proceed much in the same way as Ref.~\cite{troyer}, but we have pushed 
the simulations to higher precision, larger lattices, and lower temperatures than previously. 
The calculations are very demanding due to the fact that the impurity susceptibility is the 
difference between two extensive quantities; the total susceptibilities with and without
impurity. They are defined as
\begin{equation}
\chi_m^z = \frac{J}{T} \left ( \sum_{i=1}^{N_m} S^z_i \right )^2,
\end{equation}
with $N_0=2L^2$ when there is no impurity and $N_1=2L^2-1$ or $N_1=2L^2$ for the $S=1/2$
and $S=1$ impurities, respectively. The impurity susceptibility
$\chi_{\rm imp} = \chi_1 - \chi_0$. Very large lattices are required in order to eliminate 
finite-size effects at low temperatures---we here report results for up to $256\times 256 \times 2$
spins. We achieved relative statistical errors for $\chi_0$ and $\chi_1$ as small as 
$\approx 10^{-6}$. The use of improved estimators \cite{evertz} is crucial, but still very long 
simulations are required. The calculations reported here required approximately $5 \times 10^5$ 
Pentium III ($\approx$ 1GHz) CPU hours.

Finite-size effects are investigated by considering systems, at their quantum-critical couplings, 
of increasing size $L=4,8,16,\ldots$ at each temperature $T$, up to $L$ sufficiently large for 
ramaining finite-zise corrections to be negligible. Finally, size-converged results are 
extrapolated to zero temperature to obtain the quantum-critical Curie constant defined in 
Eq.~(\ref{eq:qcsusc});
\begin{equation}\label{eq:prefactor}
C^*=\lim _{T\to 0}T\chi _{\text{imp}}^{z}.
\end{equation}
The Curie form can be expected to apply strictly only
in the limit $T\to 0$, and in practice we have to analyze corrections to extract $C^*$. In 
Ref.~\cite{troyer}, finite-size scaled data for $T\chi _{\text{imp}}^{z}$ were linearly 
extrapolated to zero temperature. We have also found an asymptotic linear correction in all 
cases studied, but we disagree with Ref.~\cite{troyer} in regards to the temperature at which the 
linear form is valid. In the case of a vacancy in the incomplete bilayer, our results extrapolate 
clearly to a value for $C^*$ a few percent larger than $S(S+1)/3$, thus confirming an anomalous 
Curie constant (falling outside the conjectured range, however). 
For the symmetric bilayer, which is the host system previously studied by Troyer 
\cite{troyer}, the linearity of the correction to $C^*$ is established only at the lowest temperatures 
we have reached, and we cannot reliably extrapolate to obtain a direct independent confirmation of a 
universal (for given S) \cite{sachdev1} anomalous Curie constant from these results. However, 
the available low-$T$ data points are consistent with the results for the incomplete bilayer. 
For an $S=1$ impurity we also find a linear correction to the Curie form but the extrapolated 
$C^*$ is very close to the normal value. 

\begin{figure}
\includegraphics[width=7.5cm,clip]{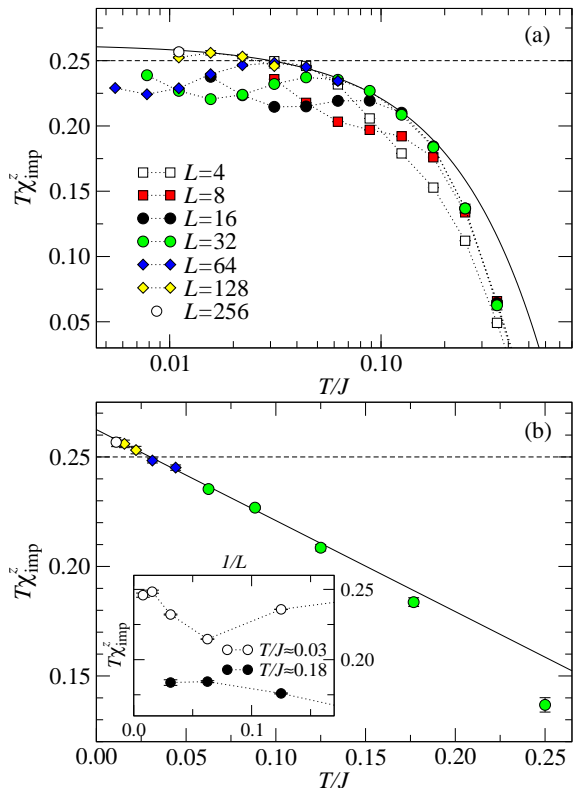}
\caption{(Color online) Impurity susceptibility multiplied by the temperature for a vacancy 
in the quantum-critical incomplete bilayer. (a) shows all our finite-size data on a log-lin 
scale and (b) shows size converged results on a lin-lin scale. Examples of finite-size 
effects are shown in the inset. The solid line in (b) and the curve in (a) show a 
linear in $T$ fit to the low-$T$ data.}
\label{fig:fig2}
\end{figure}

We now present the data underlying our conclusions summarized above.
In Fig.~\ref{fig:fig2}(a) we show the temperature dependence of $T\chi^z_{\rm imp}$ for
the incomplete bilayer, for all the system sizes we have considered and using a log-lin scale. 
Note first that as $T/J\to \infty$ the spins become independent moments exhibiting normal
Curie behavior, and thus $T\chi _{\text{imp}}^{z}\to -1/4$, due to there being one less spin in the 
system with a vacancy than in the intact system. We here focus on lower $T$. In the limit $T \to 0$, 
we have to obtain $T\chi _{\text{imp}}^{z}\to +1/4$ for any $L$ (seen in the figure only for 
$L=4$), due to the 
$S=1/2$ ground state of the even-$L$ systems with a vacancy. The temperature at which this 
can be observed is expected to scale as $1/L$, reflecting the low-energy level spacing of a 
quantum-critical system with dynamic exponent $z=1$ \cite{chn}. Interestingly, for the larger 
lattices the approach to the limiting $T=0$ value is preceded by a miniumum and a maximum. 
The finite-size behavior for a fixed low $T$ is thus also non-monotonic, as shown in the inset 
of Fig.~\ref{fig:fig1}(b). We consider $T\chi^z_{\rm imp}$ size-converged when results for 
system sizes $L$ and $L/2$ agree to within statistical errors. We have checked this 
carefully for small systems and do not see any indications of non-monotonicities beyond a
single minimum. Using this criterion we have size-converged data for temperatures down 
to $T/J = 1/64$. At $T/J=1/(64\sqrt{2})$, the infinite-size result is probably marginally
higher than our $L=256$ result. In Fig.~\ref{fig:fig2}(b) we show the size converged 
results [and the almost converged $T/J=1/(64\sqrt{2})$ data] on a lin-lin scale. A linear $T$ 
dependence can be seen below $T/J \approx 0.12$ and an extrapolation gives 
$C^* = 0.262(2)$. Note that even without an extrapolation it is 
clear that $C^*$ exceeds the normal Curie constant $C=1/4$ because the last three data points in 
Fig.~\ref{fig:fig1}(b) are all above this value. 

To check for potential systematic errors,
we have carried out extensive tests of our SSE codes, comparing results based on very long runs 
for $4\times 4$ systems with exact diagonalization data and using two independently 
written programs with different random number generators for large lattices. Recent 
calculations using a completely different method have also confirmed the absence of systematic errors 
in SSE calculations \cite{vbqmc}. Finally, we have also verified that small changes in the coupling 
$g$, of the order of the statistical errors of $g_c$ \cite{wang}, do not appreciably change 
$\chi_{\rm imp}$ in the range of tempertures studied here. We therefore consider our conclusion 
of an anomalous Curie constant to be beyond reasonable doubt.

\begin{figure}
\includegraphics[width=7.5cm,clip]{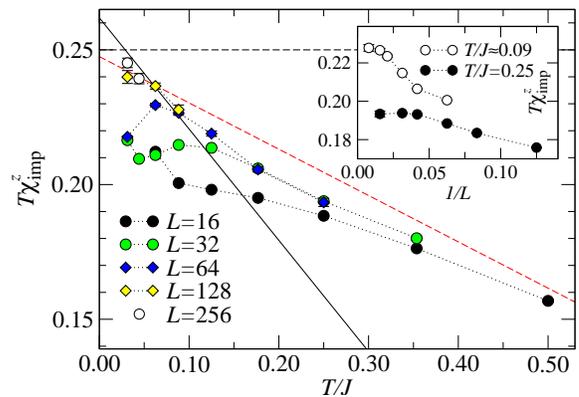}
\caption{(Color online) $T$ times the impurity susceptibility for a vacancy in the 
quantum-critical symmetric bilayer. The solid line shows the linear fit of Fig.~\ref{fig:fig2}(b). 
The dashed (red) line is the linear fit of Ref.~\cite{troyer}. The inset shows examples of the
size dependence.}
\label{fig:fig3}
\end{figure}

\begin{figure}
\includegraphics[width=7.5cm,clip]{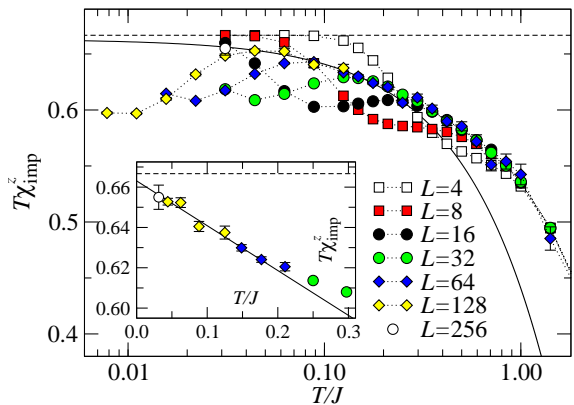}
\caption{(Color online) Impurity susceptibility of an $S=1$ vacancy in the quantum-critical 
symmetric bilayer. The inset shows size converged results and a linear fit to the
low-$T$ data.}
\label{fig:fig4}
\end{figure}

In Fig.~\ref{fig:fig3} we show data for the vacancy in a complete bilayer; the same system
that was previously studied by Troyer \cite{troyer}. The computational effort involved in 
simulating this system (for given $T,L$) is roughly twice that for the incomplete bilayer, 
because of the higher critical coupling and larger number of interactions. In addition, 
the size-convergence is slower. We have therefore not been able to go  
to as low temperatures as in the preceding case. We also find much larger corrections to the linear 
low-$T$ behavior, and as a consequence we cannot carry out a reliable $T\to 0$ extrapolation. 
However, as shown in Fig.~\ref{fig:fig3}, the line fit to the low-$T$ results for the incomplete 
bilayer in Fig.~\ref{fig:fig2} describes reasonably well also the low-$T$ results for the symmetric
bilayer. The $L=256$ result at the lowest temperature is likely not completely size converged.
The $L=256$ point at the next-lowest $T$ is also below the line, but the deviation is only two 
error bars and also here there may be some remaining finite-size effects. Moreover, the slope 
is most likely not universal and should thus depend on the low-energy parameters of the bulk 
system. The line shown may therefor not be the correct one for this system although $C^*$ 
should be universal (for given $S$) \cite{sachdev1,sachdev2}. Thus we regard these symmetric 
bilayer results compatible with the anomalous $C^*$ extracted for the incomplete 
bilayer.

In his study,  Troyer fitted a line to QMC data in the range $0.1\leq T/J\leq 0.4$ \cite{troyer}, 
resulting in the dashed line reproduced in Fig.~\ref{fig:fig3}---the $T=0$ intercept is 
consistent with $1/4$ within statistical errors. However, the line deviates 
considerably from our size-converged results for $T/J > 0.1$. The discrepancy may be due to a 
different size-convergence procedure; an $1/L$ extrapolation was mentioned in Ref.~\cite{troyer},
whereas we have used a criterion of size-independence. The latter procedure should be more
reliable because an asymptotically exponential convergence is expected at finite temperature.
An $1/L$ extrapolation can lead to a too high value when fitting only to a range of points for 
which an almost linear in $1/L$ behavior is observed, as can be seen in the inset of 
Fig.~\ref{fig:fig3} for $T/J=0.25$.

We now turn to our results for an $S=1$ impurity,  which we have realized in the symmetric 
bilayer as shown in Fig.~\ref{fig:fig1}(c). As shown in Fig.~\ref{fig:fig4}, the finite-size 
behavior of the impurity susceptibility is similar to the $S=1/2$ impurities. For $L=4,8$, 
and $16$ the asymptotic low $T$ behavior $T\chi _{\text{imp}}^{z}(T\to 0)=S(S+1)/3=2/3$ is clearly 
seen. Size converged low-$T$ results are shown in the inset. A linear behavior sets in at 
$T/J\approx 0.15$. Here the extrapolated $C^*=0.663(2)$ is slightly 
below the normal value $2/3$, but the difference is too small to definitely conclude that this is 
the case. The results do show that the anomaly is smaller than for $S=1/2$. The theory does not 
predict how the fractional spin evolves as a function of $S$ \cite{sachdev1,sachdev2}.

In summary, we have presented evidence from unbiased numerical computations
of an anomalous Curie response of an $S=1/2$ impurity spin in a 2D quantum-critical antiferromagnet.
The anomalous Curie constant $C^*=0.262(2)$ is only $\approx 5\%$ larger than the normal $C=1/4$.
For $S=1$ we obtain $C^*$ marginally below the normal value, $2/3$, but better statistics is needed
to confirm this.

It should be noted that the anomalous Curie constant cannot be strictly interpreted as due to a 
fractionalized impurity spin, because it is a finite-temperature quantity with contributions 
from many states with different total spin, even as $T \to 0$. Fractionalization does not occur 
in the ground state for finite $L$, but an interesting universal impurity-induced spatial structure 
has been found \cite{hss}. Recently impurity effects have been examined theoretically also in
fractionalized spin liquid states \cite{kolezhuk,florens}. 

We would like to thank S.~Sachdev, M.~Troyer, and M.~Vojta for useful discussions. 
KHH acknowledges a travel grant awarded by the Finnish Academy of Science and Letters from the 
Vilho, Yrj\"{o}, and Kalle V\"{a}is\"{a}l\"{a} Foundation. AWS is supported by the NSF under Grant 
No.~DMR-0513930. Part of the simulations were performed at the CSC, the Finnish IT Center for Science.

\null


\begin{thebibliography}{00}
\bibitem{note0}
The criticism was met by Sachdev and Vojta in Ref.~\cite{sachdev2}.
\end{thebibliography}
\end{document}